\documentclass{nature}
\usepackage{psfig}

\title{A dark jet dominates the power output of the stellar black hole
Cygnus X-1} 

\author{Elena Gallo$^1$,$~~$Rob Fender$^{1,2}$,$~~$Christian Kaiser$^2$,
  $~~$David Russell$^2$,$~~$Raffaella Morganti$^3$,\\
    Tom Oosterloo$^3$,$~~$Sebastian Heinz$^4$}

\long\def\symbolfootnote[#1]#2{\begingroup%
\def\thefootnote{\fnsymbol{footnote}}\footnote[#1]{#2}\endgroup} 
\def\blfootnote{\xdef\@thefnmark{}\@footnotetext}
 
\def\gtsima{$\; \buildrel > \over \sim \;$}
\def\ltsima{$\; \buildrel < \over \sim \;$}
\def\gsim{\lower.5ex\hbox{\gtsima}}
\def\lsim{\lower.5ex\hbox{\ltsima}}
\def\simgt{\lower.5ex\hbox{\gtsima}}
\def\simlt{\lower.5ex\hbox{\ltsima}}
\def\simpr{\lower.5ex\hbox{\prosima}}

\begin{document}

\maketitle
\small
\begin{affiliations}

\item Astronomical Institute `Anton Pannekoek' and Center for High Energy
Astrophysics, University of Amsterdam, Kruislaan 403, 1098 SJ Amsterdam, NL
\item School of Physics and Astronomy, University of Southampton,
Highfield, Southampton, SO17 1BJ, UK
\item  Netherlands Foundation for Research in Astronomy, Postbus 2,
NL-7990 AA, Dwingeloo, NL
\item  Center for Space Research, Massachusetts Institute of
Technology, 77 Massachusetts Avenue, Cambridge, MA 02139, USA

\end{affiliations}
\normalsize

\begin{abstract}
Accreting black holes are thought to emit the bulk of their power in the X-ray
band by releasing the gravitational potential energy of the infalling
matter\cite{fkr}. At the same time, they are capable of producing highly collimated jets
of energy and particles flowing out of the system with relativistic
velocities\cite{hughes}. Here we show that the 10 solar mass black hole in the
X-ray binary Cygnus
X-1\cite{bowyer}$^{,}$\cite{giesbolton}$^{,}$\cite{herrero} is surrounded by a
large-scale (about 5 pc in diameter) ring-like structure that 
appears to be inflated by the inner radio jet\cite{stirling}. We estimate that
in order to sustain the observed emission of the ring, the jet of Cygnus X-1
has to carry a kinetic power that can be as high as the bolometric X-ray
luminosity of the binary system.  This result may imply that low-luminosity
stellar mass black holes as a whole dissipate the bulk of the liberated
accretion power in the form of `dark', radiatively inefficient relativistic
outflows, rather than locally in the X-ray emitting inflow.
\end{abstract}
\normalsize
Relativistic jets are a common feature of accreting black holes on all mass
scales, from super-massive black holes at the centres of active galactic
nuclei\cite{urry}$^,$\cite{blandford} to stellar mass black holes in X-ray
binary systems within our own
Galaxy\cite{mr94}$^,$\cite{mr99}$^,$\cite{fender05}.  While the inflow of hot
gas can be very efficient in producing light (up to $\sim 40$ per
cent of the accreted material may be transformed into energy and radiated
away in the form of optical/UV/X-ray photons) the same is not true for for the 
synchrotron-emitting outflow, whose efficiency might be lower than a few per
cent. Estimating the {\it total} -- radiated plus kinetic -- power content
of the jets, and hence their importance with respect to the accretion process
in terms of energetics,
is a primary aim of high energy astrophysics.

We observed the field of the 10 solar mass black hole and Galactic jet source
Cygnus X-1 at 1.4 GHz for 60 hours with the Westerbork Synthesis Radio
Telescope (WSRT), yielding the deepest low frequency radio observation of that
region to date\cite{martietal96}.  A ring of radio emission -- with a diameter
of $\sim$1 million AU -- appears northeast of Cygnus X-1 (Figure 1), and seems
to draw an edge between the tail of the nearby HII nebula
Sh2-101\cite{sharpless} (whose distance is consistent with that to Cygnus
X-1\cite{hunter}) and the direction of the radio jet powered by Cygnus
X-1\cite{stirling}. Since Cygnus
X-1 moves in the sky along a trajectory which is roughly
perpendicular to the
jet\cite{lestrade}$^{,}$\cite{stirling}$^{,}$\cite{mir03}, and thus can not 
possibly be traced back to the ring centre, this rules out that the ring might
be the low-luminosity remnant of the natal supernova of the black hole.  In
analogy with extragalactic jet sources, the ring of Cygnus X-1 could be the
result of a strong shock that develops at the location where the collimated
jet impacts on the ambient interstellar medium (Figure 2). The jet particles
inflate a radio lobe which is over-pressured with respect to the surroundings,
thus the lobe expands sideways forming a spherical bubble of shock-compressed
ISM, which we observe as a ring because of limb brightening effects. The
collisionally ionized gas behind the bow shock would produce the observed
bremsstrahlung radiation; in addition, if the shock is radiative, significant
line emission is expected from hotter gas at the bow shock front.  Structures
similar to the ring of Cygnus X-1 have been found at the edges of radio
lobes inflated by the jets of super-massive black
holes at the centre of powerful radio galaxies\cite{smith}, where the much
higher temperatures of the intra-cluster 
medium compared to the ISM shift the bremsstrahlung emission to X-ray
frequencies.  Striking confirmation of this interpretation comes from
follow-up optical observations of the field of Cygnus X-1 with the Isaac
Newton Telescope Wide Field Camera: the ring is clearly detected using a
H$\alpha$ filter in an exposure of only 1200 sec (Figure 3). The estimated
flux of the ring at H$\alpha$ frequencies exceeds the measured
radio flux by a factor $\simgt$20, indicating that the collisionally
ionized gas in the ring is indeed emitting bremsstrahlung radiation and also
that a significant amount of the measured H$\alpha$ flux is due to line
emission, as expected in the case of radiative shock.

Acting as an effective jet calorimeter, the ISM allows an estimate of the
jet's {\em power$\times$lifetime} product\cite{burbidge} that is, in
principle, independent of the uncertainties associated with the jet spectrum
and radiative efficiency. Following a self-similar fluid model developed for
extragalactic jet sources\cite{castor}$^{,}$\cite{ka97}$^{,}$\cite{hrb}, we
assume that the 
jet of Cygnus X-1 is supplying energy at a constant rate $P_{\rm jet}$, and is
expanding in a medium of constant density. We set the minimum temperature of
the thermal gas to $T_{\rm shock} \simgt 10^4$ K, a typical temperature above
which the cooling time becomes critically short, and below which the
ionization fraction becomes too low for the ring to emit observable
bremsstrahlung radiation.  Given the average ring monochromatic luminosity, we
are able to estimate a density of about $1300$ cm$^{-3}$ for the
shock-compressed particles in the ring from the expression for the bremsstrahlung
emissivity\cite{longair}$^{,}$\cite{lotz}.  By balancing the interior
pressure exerted by the lobe and the ram pressure of the shocked ISM, it can
be shown\cite{ka97} that the jet length within the lobe grows with the time
$t$ in such a way that: $ t
\simeq (L/2)^{(5/3)}\times(\rho_0 / P_{\rm jet})^{(1/3)}$, being $L$ the
separation between Cygnus X-1 and the ring's outermost point, and 
$\rho_0$ the mass density of the un-shocked gas. By writing the time
derivative of this equation, there follows a simple relation between the jet
lifetime, its length within the lobe, and the ring velocity: $t=\frac{3}{5}(L
/ \rm v_{\rm ring})$.
For a strong shock in a mono-atomic gas, the expansion velocity is set by the
temperature of the shocked gas. If the shock is radiative, then the initial
post-shock temperature can be higher than that of the thermalized,
bremsstrahlung-emitting gas. A
stringent constraint comes from X-ray 
observations: from the non-detection of soft X-ray emission in a 12 ksec
observation of Cygnus X-1 taken with the Chandra X-ray Observatory, we can place an upper
limit of $~T_{\rm shock}\simlt3\times 10^6~$ K by modelling the emission as a
radiative shock. 
This, combined with the lower limit of $10^4$ K, gives a ring velocity $~\rm
v_{\rm ring}\simeq 20-360$ km sec$^{-1}$.
The resulting jet lifetime is $~t\simeq 0.02-0.32$ Myr, to be compared
with the estimated age of the 
progenitor of the black hole in Cygnus X-1, of a few Myr\cite{mir03}. 

Adopting a mass density of the un-shocked gas that is 4 times lower than the initial
post-shock density, we infer a {\em
~time-averaged~} energy emission rate from the jet between $8\times 10^{35}$
and $10^{37}$ erg 
sec$^{-1}$. Based upon daily X-ray and radio monitoring of Cygnus
X-1 over the last 10 years, we know that Cygnus X-1 is in a hard X-ray
state\cite{mccr} 
-- and hence powers a collimated jet\cite{fender05} -- for about 90 per cent of its
lifetime. Taking this duty cycle into account, the total power carried by the
jet of Cygnus X-1 is: $~9\times 10^{35} \simlt P_{\rm jet}\simlt 10^{37}$ erg
sec$^{-1}$, up to two orders of magnitude higher than the existing estimate
based on the flat radio spectrum\cite{fenderetal01}. The fact that the jet of
Cygnus X-1 switches off for short periods of times (typically for a few months
over timescales of years) does not violate the model assumptions: the
condition of constant power supply is met as long as the jet is intermittent
over timescales that are short compared to its lifetime.
The total power carried by the jet is a significant fraction
$f\simeq 0.03-0.5$ of the bolometric (0.1-200 keV) X-ray luminosity of
Cygnus X-1 while in the hard state\cite{ds01}; the total energy deposited by
the jet into the surrounding ISM over its lifetime is $\simeq 7\times 10^{48}$
erg.

%
The particle density of the ISM through which the ring is expanding is
constrained between $\sim 1$ and about 300 cm$^{-3}$, at most three orders of
magnitude higher than the average ISM density in the Galaxy. The lack of a
`counter-ring' can be explain in terms of a much lower particle density in the
opposite direction to Cygnus X-1.  Such large density inhomogeneities are not
unusual for dense star forming regions, such as the Cygnus association, and
would support the hypothesis that the ring is the result of the interaction
between the radio lobe and the tail of the HII nebula. If so, the
counter-jet of Cygnus X-1 is travelling undisturbed to much larger distances,
gradually expanding and releasing its enormous kinetic energy. This could mean
that the ring of Cygnus X-1 is a rather exceptional detection for this class
of objects, made possible by its proximity to the HII nebula. 
Taking into account the contribution of the counter-jet as well, the total power
dissipated by the jets of Cygnus X-1 in the form of kinetic 
energy can be as high as the bolometric X-ray luminosity of the system ($f=0.06-1$).

The results presented here have important consequences for low-luminosity
stellar mass black holes {\em as a whole}: several
works\cite{gfp}$^,$\cite{lpk}$^,$\cite{mmf} have suggested that hard state
stellar black holes below a critical X-ray luminosity dissipate most of the
liberated gravitational power in the form of radiatively inefficient outflows,
rather than locally in the accretion flow. This is because in hard state black
hole binaries the total jet power and the observed X-ray luminosity follow a
non-linear relation\cite{fgj} of the form $P_{\rm jet}\propto L_{\rm
X}^{~~0.5}$ (both expressed in Eddington units). Using the ring of Cygnus X-1
as an effective 
calorimeter for the jet power, we have constrained the normalization factor of
the above equation, showing that $P_{\rm jet}= fL_{\rm X}$, with $f\simeq
0.06-1$, when $L_{\rm X}\simeq 0.02$. Thus the critical X-ray luminosity
below which $P_{\rm jet}> L_{\rm X}$ is {\em~no lower~} than a few $10^{-5}$
Eddington, and could even be as high as the peak luminosity of the hard X-ray
state. This radically alters our concepts of the accretion process and of the
feedback of accretion power into the surroundings. Via the new observations 
presented here we have strong evidence that the power output of low-luminosity 
-- i.e. the overwhelming majority of 
-- stellar black holes is dominated by the kinetic energy of `dark'
outflows, whose key signature is the eventual energisation of the ambient
medium.

\noindent
Correspondence and requests for materials should be addressed to
E.G. (egallo@science.uva.nl)  

\noindent
We are grateful to Dimitris Mislis and Romano Corradi for the H$\alpha$
observation presented in this work. The Westerbork Synthesis Radio Telescope
is operated by the ASTRON (Netherlands Foundation for Research in Astronomy)
with support from the Netherlands Foundation for Scientific Research (NWO).
\newpage







\begin{figure*}
\vspace{-2cm}
\hspace{-1cm}
\centerline{\psfig{figure=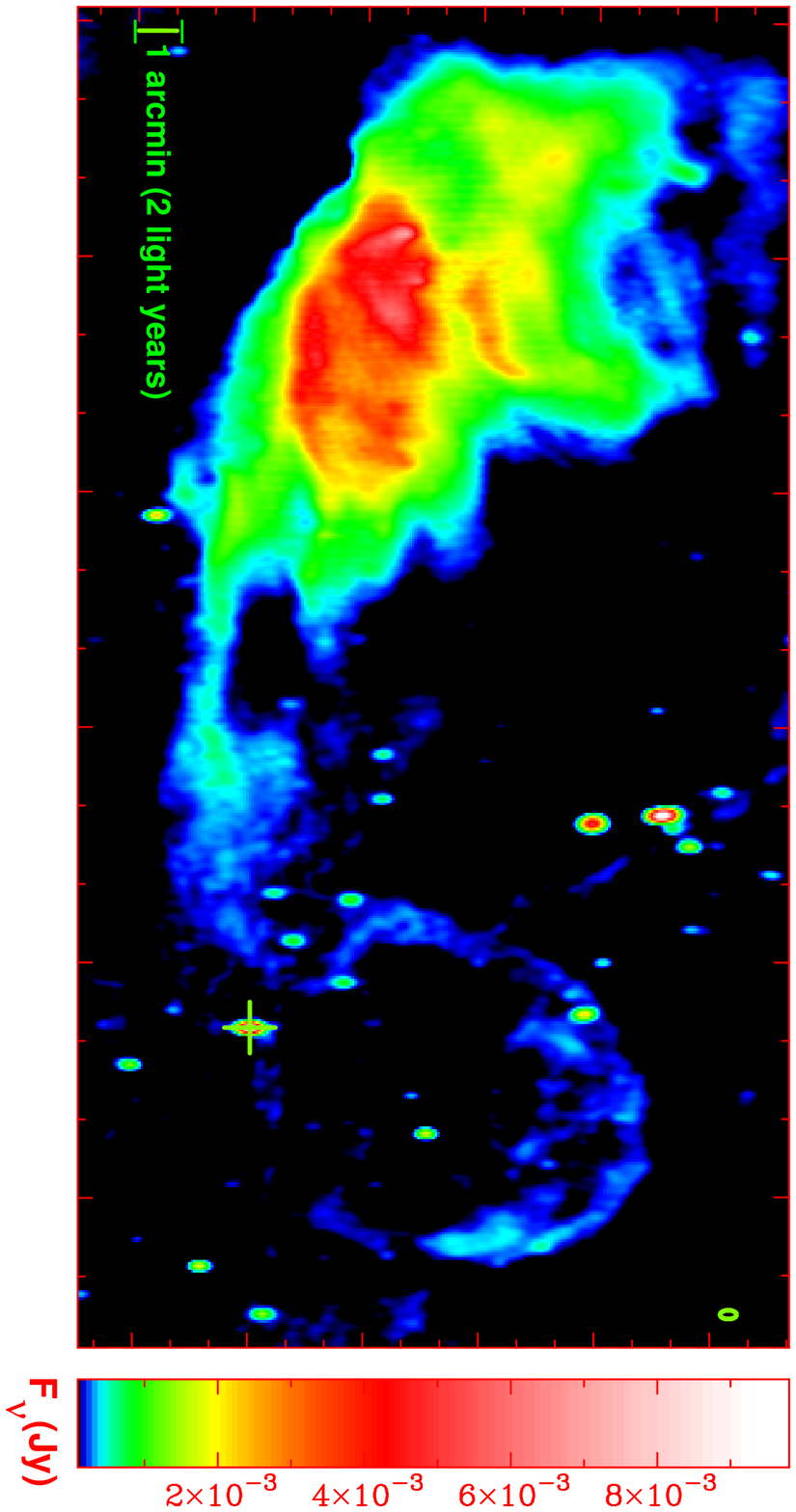,angle=180,width=18cm}}
\end{figure*}
\begin{figure*}
\caption{\small \sf{The interstellar gas around a Galactic stellar black
hole is stirred by the pressure of a highly collimated
relativistic jet of energy and particles, resulting in a 15 light-years wide
ring of radio emission.  The field of view of the 10 solar mass black hole
in Cygnus X-1 (marked by a cross) 
was observed by the Westerbork Synthesis 
Radio Telescope for 60 hours at 1.4 GHz: the ring appears
to draw an edge between the tail of Sh2-101, the nearby HII nebula on the left
hand side, and the direction of the inner radio jet of Cygnus X-1 (shown in
the inset of Figure
2). The spatial resolution in this map ($25\times 14$ arcsec$^2$) is
illustrated by the green open ellipse on the top right corner. The wedge shows
the (logarithmic) flux scale, between 0--0.01 Jy; the average ring
monochromatic flux is 0.2 mJy beam$^{-1}$. At a distance of 2.1 kpc, the
separation $L$~between Cygnus X-1 (coincident with the jet base) and the
ring's outermost point is $1.9\times 10^{19}\times$sin($\theta$)$^{-1}$ cm,
where $\theta$ is the jet inclination to the line of sight ($\theta \simeq
35^{\circ}$$^[$\cite{herrero}$^]$). 
Because of limb-brightening, we observe a ring 
whose thickness $\Delta R$~in the plane of the sky equals the effective length
we are looking through into the bubble. At 2.1 kpc, $\Delta
R\simeq 1.6\times 10^{18}$ cm.}}
\end{figure*}
\clearpage

%
\begin{figure*}
\hspace{-1cm}
\centerline{\psfig{figure=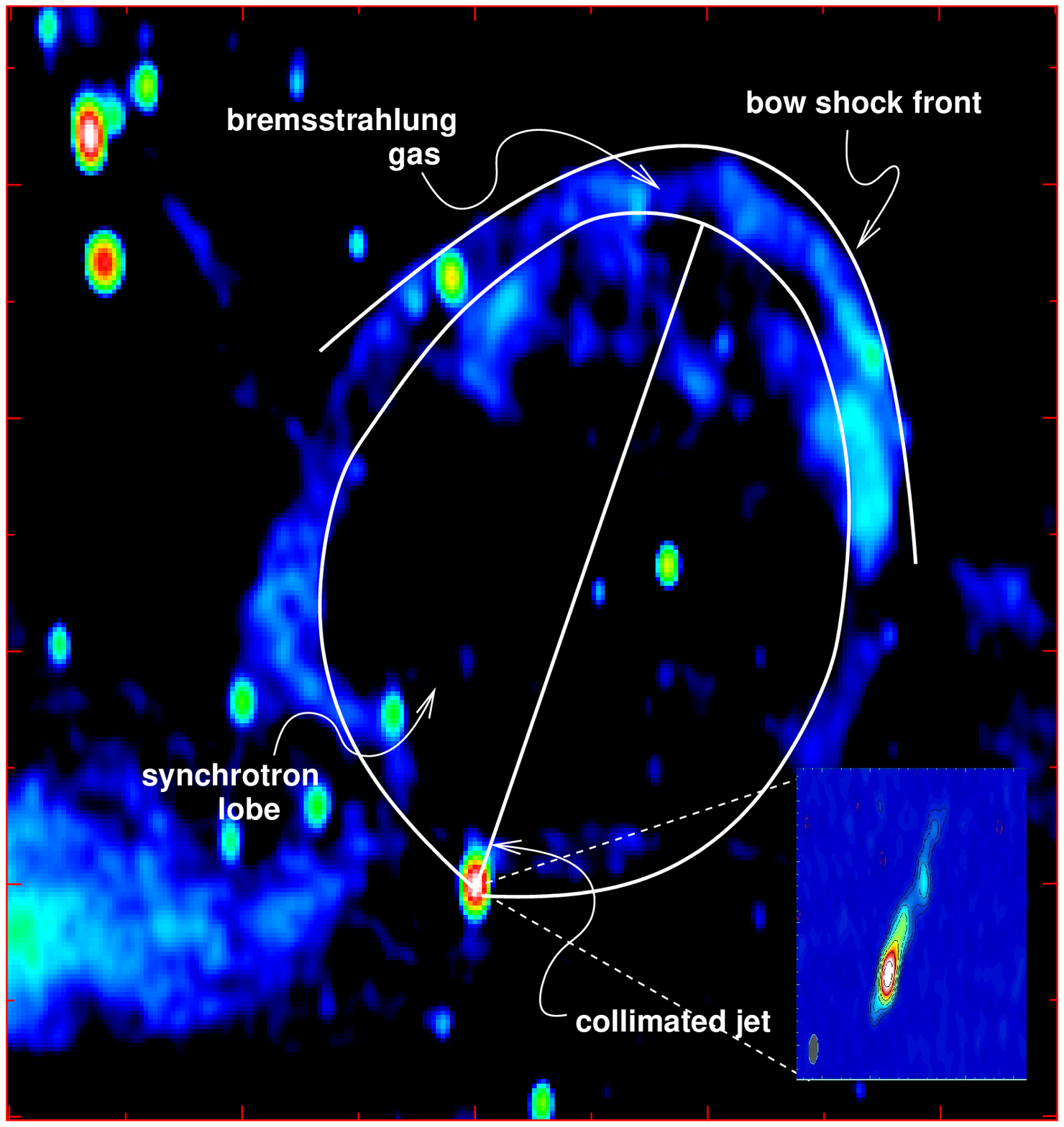,angle=-90,width=21cm}}
\end{figure*}
\clearpage
\begin{figure*}
\caption{\small \sf{
The ring of Cygnus X-1 is the result of a strong shock that develops at the
location where the pressure exerted by the collimated milliarcsec-scale jet,
shown in the inset, is
balanced by the interstellar medium.  The jet particles start to inflate a
synchrotron-emitting lobe which is over-pressured with respect to the
surrounding gas, thus the lobe expands sideways forming a spherical bubble of
shock-compressed bremsstrahlung-emitting gas. The
monochromatic luminosity of the ring, $L_{1.4~\rm GHz}\simeq
10^{18}$ erg sec$^{-1}$ Hz$^{-1}$, equals the product
($\epsilon_{\nu}\times V$), where the source unit volume $V$ is given by the beam
area times the measured ring thickness: $V
\simeq 4 \times 10^{53}$ cm$^3$, and $\epsilon_{\nu}$ is the expression of the
bremsstrahlung emissivity for a pure  hydrogen gas emitting at a temperature $T$:
$\epsilon_{\nu}=6.8\times 10^{-38}~ g(\nu,T) 
~T^{(-1/2)}~ n_e^2~ {\rm exp} (h\nu/k_B T)$~erg cm$^{-3}$ sec$^{-1}$ Hz$^{-1}
$ (being $h$ and $k_B$ the Plank and Boltzmann's constant, respectively).     
For $T\simeq 10^4$ K and a Gaunt factor 
$g\simeq 6$, the density $n_e$ of the ionized particles in the ring is 
$n_e \simeq 25$   
cm$^{-3}$. The ionization fraction at $10^4$ K~is
$x\simeq0.02$$^[$\cite{lotz}$^]$,  
resulting in a total
particle density $n_t\simeq 1300$ cm$^{-3}$. 
The minimum pressure inside the lobe predicted by the model is $\simeq 5\times
10^{-11}$ erg cm$^{-3}$.  If this pressure is solely due to a magnetized
relativistic pair plasma in equipartition, then the strength of the magnetic
field is about 40 $\mu$G. Assuming minimum energy conditions\cite{longair},
this yields an expected lobe synchrotron surface brightness of $\simeq$ 35 mJy
beam$^{-1}$ at 1.4 GHz, more than 150 times brighter than the observed
ring. 
The much lower upper limit
for the lobe surface brightness means that either
the system is far from equipartition, or the most of the energy is stored in
non-radiating particles, presumably baryons.
}}
\end{figure*} 

\clearpage


\begin{figure*}
\hspace{-1cm}
\centerline{\psfig{figure=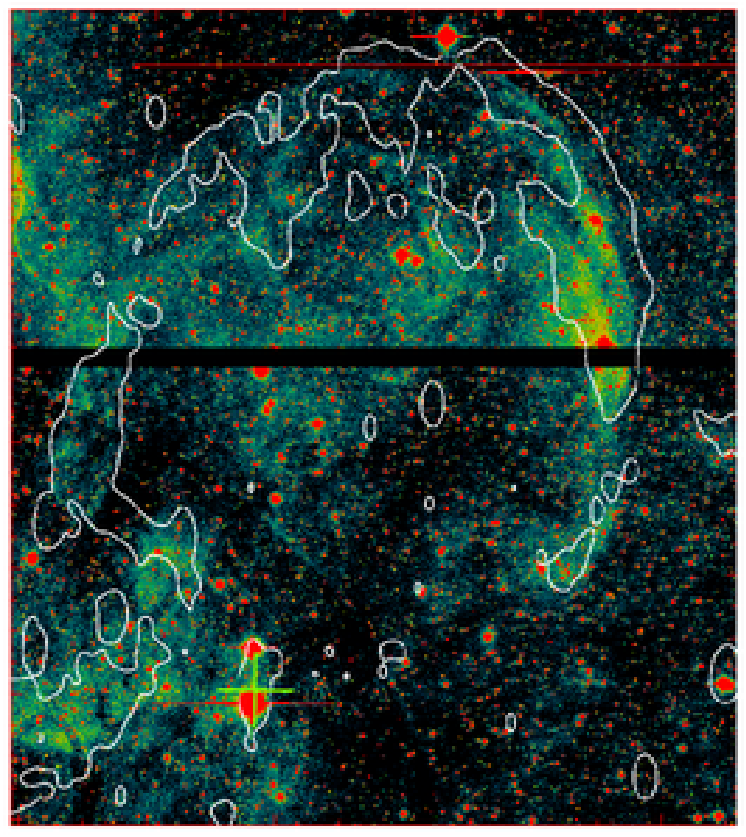,angle=270,width=18cm}}
\end{figure*}
\begin{figure*}
\begin{center}
\caption{\small \sf{Optical counter-part of the radio ring of
Cygnus X-1. The optical image, taken with the Isaac Newton
Telescope Wide Field Camera using an H$\alpha$ filter, is shown with the
3$\sigma$ radio contours over-plotted in white. As no calibration was taken
during the observation, an absolute flux scale can not be set;  however, given that the ring is clearly
detected in a 1200 sec exposure, and taking into account the atmospheric and
sky conditions during the observation, this translates into a minimum
unabsorbed H$\alpha$ flux of 0.02 mJy arcsec$^{-2}$. The corresponding
radio-optical spectral index $a$ (defined such that $F_{\nu}\propto \nu^a$) is
greater than 0.2. This implies an emission mechanism with flat spectrum, such
as bremsstrahlung, plus excess flux possibly due to line emission, as expected
in the case of radiative shock. For
comparison, if the ring emitted optically thin synchrotron radiation with a
spectral index $a=-0.7$, the expected flux at H$\alpha$ frequencies would be
1.2 10$^{-7}$ mJy arcsec$^{-2}$, by no means detectable by the INT WFC in 1200
sec.  }  }
\end{center}
\end{figure*}
\clearpage

\normalsize
\end{document}